\documentclass[reprint,aps,prc,twocolumn,superscriptaddress]{revtex4-2}
\usepackage{amsmath}
\usepackage{amssymb}
\usepackage{nccmath}
\usepackage{bm}
\usepackage{float}
\usepackage{slashed}
\usepackage{comment}
\usepackage{graphicx}
\usepackage{adjustbox}
\usepackage{rotating}
\usepackage{braket}
\usepackage{ulem}
\usepackage{placeins}
\usepackage{longtable}

\date{}
\newcommand\T{\rule{0pt}{3ex}}       
\newcommand\B{\rule[-1.5ex]{0pt}{0pt}} 
\newcommand{\colorcaption}[2][]{%
	\begingroup%
	\renewcommand{\@caption@fignum@sep}{ (Color online). }%
	\caption[#1]{#2}%
	\endgroup%
}

\usepackage{hyperref}
\hypersetup{
	colorlinks=true, 
	linktoc=all,     
	linkcolor=blue,  
	urlcolor=blue,
}

\begin{document}
		\large
\title{Evaluation of bound-state \texorpdfstring{$\beta^-$}{beta-}-decay half-lives of fully ionized atoms}

\author{Priyanka Choudhary\footnote{pricho@kth.se}}
\email{pricho@kth.se} 
\author{Chong Qi\footnote{chongq@kth.se}}
\email{chongq@kth.se} 
\address{Department of Physics, KTH Royal Institute of Technology, Roslagstullsbacken 21, SE-106 91 Stockholm, Sweden}

\date{\hfill \today}


\begin{abstract}Bound-state $\beta^-$-decay is a rare radioactive process where the created electron is trapped in an atomic orbital instead of being emitted. It can be observed in highly ionized atoms in particular when normal beta decay is energetically forbidden, but bound-state decay is still possible.
In this work we present a systematic theoretical study on the bound-state $\beta^-$-decay of fully ionized atoms where key nuclear inputs include the nuclear matrix elements (expressed through $ft$ values) and the lepton phase-space volume function. We present a method to evaluate nuclear matrix elements for fully forbidden $\beta^-$ transitions in neutral atoms, from the inverse electron capture process using the Takahashi–Yokoi model and account for the impact of electron capture to different atomic orbitals on the resulting half-lives. Decay rates for bound-state $\beta^-$-decays of nuclei $^{163}$Dy, $^{193}$Ir, $^{194}$Au, $^{202}$Tl, $^{205}$Tl, $^{215}$At, $^{222}$Rn, $^{243}$Am, and $^{246}$Bk are calculated, where the normal beta decay is forbidden. In addition, we compute the bound-state $\beta^-$ decay rates for nuclei $^{187}$Re, $^{227}$Ac, and $^{228}$Ra, observing enhancements by factors of $10^2$ to $10^4$ relative to their neutral-atom counterparts. Our results show that the half-lives of certain bare nuclei are significantly shorter than those of the corresponding neutral atoms, identifying them as promising candidates for future experimental investigation. The theoretically predicted half-lives of the bound-state $\beta^-$ decay could provide valuable inputs for various astrophysical studies. 

\end{abstract}


\maketitle

\section{Introduction}
Nuclear ${\beta}$ decay, governed by the weak nuclear force, is one of the earliest known fundamental processes of radioactive decay. This process is crucial for understanding the stability of nuclear matter \cite{EPJA}, the nucleosynthesis of elements~\cite{RevModPhys.49.77,Langanke_2021}, as well as for understanding the weak interaction and the nature of many fundamental particles including the neutrinos~\cite{Hayen}.
${\beta}$ decay involves several modes: beta minus (${\beta}^-$) decay, beta plus (${\beta}^+$) decay, electron capture (EC), as well as the rare neutrino capture.

The EC of the nucleus (of an electrically neutral atom) undergo mostly by absorbing one of its own inner atomic electrons (usually from the $K$ or $L$ shell). A fully/completely ionized nucleus (often called bare atom) can capture a free electron from its surroundings. Such process can be expected in high-temperature environments like collapsing cores of supernovae, where atoms can lose all their electrons through ionization due to their intense thermal energy. The ${\beta}^-$ decay can be considered as the mirror process of the EC. The difference, however,  is that  ${\beta}^-$ decay mostly emits high-energy electrons into the continuum, as the decay $Q_n$ value is often much larger than the electron binding energy. Bound-state ${\beta}^-$-decay ($\beta_b$-decay), first proposed by Daudel \textit{et al.} \cite{163Dy}, is a process in which the emitted electron occupies a vacant bound atomic orbital instead of escaping into the continuum \cite{Litvinov_2011}.  While this channel can be negligible for most neutral atoms due to the limited available phase space in outer shells, it can become important in highly ionized atoms where more electron orbitals are vacant.
That could be the case again in stellar plasma characterized by extreme temperatures and densities, where atoms are often partially or fully ionized. That high temperature environment can provide an ideal ground for the $\beta_b$-decay to dominate. In other words, the $\beta_b$-decay offers a unique probe of how electron environments influence nuclear weak processes, particularly in extreme astrophysical conditions like supernovae or neutron star mergers.

Nuclear ${\beta}$-decay plays a pivotal role in the rapid neutron-capture process ($r$-process) \cite{RevModPhys.93.015002}, slow neutron-capture process ($s$-process) \cite{RevModPhys.83.157}, and the rapid proton-capture process ($rp$-process) \cite{SCHATZ1998167}. In particular, the $s$-process is responsible for synthesizing approximately half of the elements heavier than iron and proceeds via a sequence of neutron captures and ${\beta}^-$-decays, typically involving low-lying states in nuclei near the valley of stability. Notably, in stellar environments, ${\beta}$-decay rates can be significantly altered by incorporating $\beta_b$-decay channels, which become prominent in highly ionized conditions. An important nucleus in this context is $^{205}$Pb, produced via the $s$-process. It can play a significant role in understanding early solar system chronology, as its decay products have been observed in ancient meteorites. However, uncertainties in the weak decay rates of $^{205}$Tl and $^{205}$Pb under stellar conditions have limited our ability to predict $^{205}$Pb abundance \cite{refId01}. Recently, a breakthrough experiment at GSI measured, for the first time, the $\beta_b$-decay of fully ionized $^{205}$Tl$^{81+}$ using a storage ring \cite{Leckenby2024,PhysRevLett.133.232701,EurPhysJA.61.130,leckenby2025bayesianmontecarloapproaches}. Another well-known example is the beta decay of $^{187}$Re, whose half-life decreases from 
$42$ Gyr in the neutral atom to just 32.9 years in the fully ionized state due to the bound-state decay channel to $^{187}$Os  \cite{PhysRevLett.77.5190}. 

In this paper we would like to present a systematic study on the $\beta_b$-decay half-lives of fully ionized atoms.
We are particularly interested in investigating transitions where $Q_n$-values are negative or very small (in the neutral atom), rendering the ${\beta}^-$-decay energetically forbidden or hindered.  In the former case, the only possible decay mode is the $\beta_b$-decay in the fully ionized (bare) atom. These transitions include the decay of $^{163}$Dy, $^{193}$Ir, $^{194}$Au, $^{202}$Tl, $^{205}$Tl, $^{215}$At, $^{222}$Rn, $^{243}$Am, and $^{246}$Bk. Latter transitions involve nuclei $^{187}$Re, $^{227}$Ac, and $^{228}$Ra, for which ${\beta}^-$-decay occurs in both neutral and bare atoms. Among these nuclei, the experimentally measured half-lives are available for the decay of  $^{163}$Dy \cite{PhysRevLett.69.2164}, $^{187}$Re \cite{PhysRevLett.77.5190}, and $^{205}$Tl \cite{Leckenby2024}. Our aim is to predict the $\beta_b$-decay half-lives or decay rates using the experimental information from the analogous inverse EC process or beta decay and the recent AME2020 atomic mass data or $Q_n$-values \cite{Wang_2021,Kondev_2021}. We hope this systematic research can contribute to our understanding of weak interaction processes in stellar environments and their impact on nucleosynthesis and nuclear chronometry.

This paper is organized as follows: In Section \ref{Methodology}, we describe the theoretical framework employed to calculate the decay rates of $\beta_b$-decay in bare atoms. Section \ref{Results} presents the results of the calculated half-lives and decay rates for the selected nuclear transitions. Finally, Section \ref{Summary} provides a summary of the study.

\section{Methodology}\label{Methodology}

\begin{figure}
    \centering
    \includegraphics[width=0.5\textwidth]{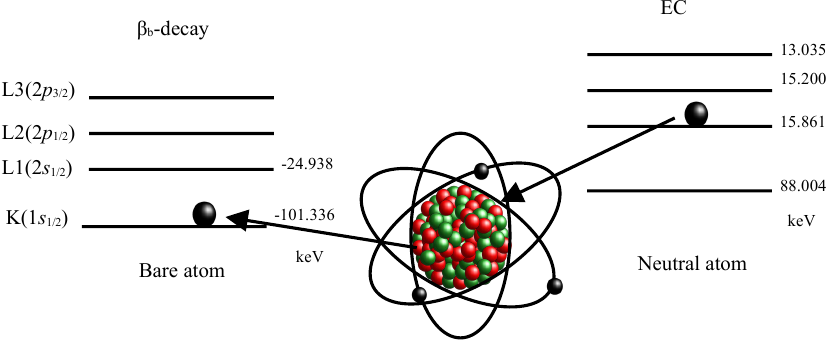}
    \caption{Illustration of the $\beta_b$ decay of the bare atom and the analogous inverse EC process of the neutron atom. The energies of the electron orbitals of the bare atom would be different from that of the neutral atom. In the latter case, the electron orbitals are affected by the Coulomb repulsion among the electrons. As examples, we listed the energies of the lowest few states for Pb bare atom and neutral atom. For the atoms of concern in this paper, most EC processes capture electrons from orbital L and those above, while the emitted electron from the bare atom may be trapped in the lowest K orbital as well as higher orbitals.}
    \label{fig:enter-label}
\end{figure}

We consider the $\beta_b$-decay as the inverse process of the EC from a nuclear matrix element perspective, as illustrated in Fig. \ref{fig:enter-label}.
The $\beta_b$-decay rates between certain initial and final states can be evaluated as 
\begin{equation}\label{Eq.1}
    \lambda = \Big(\frac{\rm ln \, 2}{ft}\big)\times f_m^*,
\end{equation}
where $m$ represents the mode of decay such as allowed, unique, and non-unique forbidden transitions, and $f_m^*$ is the lepton phase volume part. The nuclear matrix elements, which can be difficult to evaluate theoretically, can be extracted from the experimental $ft$ value from either the inverse EC process or beta decay in the neutral atom, depending on the decay energy.  
Takahashi and Yokoi developed a theoretical formalism to investigate both continuum- and bound-state ${\beta}$-decays for bare atoms \cite{TAKAHASHI1983578,TAKAHASHI1987375,PhysRevC.36.1522}.
In those works, the experimental log$ft$ values for EC where measured, and estimated values from the neighboring transitions with the same spin-parity for unobserved transitions are used to determine the $\beta_b$-decay half-life. Liu \textit{et al.} recently applied that method to study the $\beta_b-decay$ of highly ionized atoms \cite{PhysRevC.104.024304,Liu_2022} and found that the $\beta_b$-decay half-lives in the bare atoms are significantly different than those in the neutral case. To calculate the half-lives for the $\beta_b$-decay with a negative neutral-atom $Q_n$-value (the former case), as in Ref. \cite{TAKAHASHI1983578} and recently in Refs. \cite{PhysRevC.104.024304,PhysRevC.100.064313}, we begin with the experimental half-lives for the EC process and then extract the nuclear matrix elements using the phase space factor. We also determine the contribution from different electron orbitals to quantify their impact on the resulting half-lives. For transitions where experimental EC half-lives are not available, we determine the decay rate ratio of $\beta_b$-decay and EC. For transitions with a positive $Q_n$ value, as in Ref. \cite{PhysRevC.100.064313}, we adopt the experimental $\beta^-$ decay half-lives of neutral atoms for calculating the bound-state decay rates. 

For the $\beta_b$-decay, $f_m^*$ is given by
\begin{equation}\label{Eq.2}
    f_{{m}}^* = \sum_x \sigma_x (\pi/{2}) [f_x \, {\rm or} \, {\rm g}_x]^2 q^2 S_{(m)x}
\end{equation}
where, $\sigma_x$ represents the vacancy of the electron orbit $x$. We have taken it as unity in our calculations \cite{PhysRevC.100.064313}, which is expected for a fully ionized or bare atom. The quantity $q$ is given by $ q = Q_{\beta_b} /m_ec^2$. $Q_{\beta_b}$ denotes the decay energy and is described as
$$Q_{\beta_b} = Q_n - \Delta B_e + B_{K,L,...},$$
where $Q_n$-value is the mass difference between parent and daughter atoms in the neutral case. In this work, we have taken the $Q_n$-values from the new AME2020 atomic mass data \cite{Wang_2021,Kondev_2021}. $\Delta B_e$ is the difference between the total binding energies for bound electrons of the neutral daughter and parent atoms $[B_n(Z) -B_n(Z-1)]$ \cite{RODRIGUES2004117}. $B_{K,L,...}$ is the ionization energy of the different atomic orbital of the bare daughter atom \cite{NIST}.

 The $S_{(m)x}$ is the spectral shape factors, which is defined as
\begin{equation}\label{Eq.3}
    S_{(m)x} = 
    \begin{cases}
  1      & \text{for $m = a, nu$ and $x = ns_{1/2}$, $np_{1/2}$} \\
  q^2 & \text{for $m = u$ and $x = ns_{1/2}$, $np_{1/2}$} \\
  9/R^2 & \text{for $m = u$ and $x = np_{3/2}$, $nd_{3/2}$} \\
  0 & \text{otherwise}.
\end{cases}
\end{equation}

The Coulomb amplitudes for the bound electron wave functions $[f_x \, {\rm or \, g}_x]$ are defined in the following way
\begin{align}
\label{Eq.4}
    & f_x = \left(\frac{\lambdabar_c}{a_0}\right) ^\frac{3}{2} \frac{P(R)}{R}, \nonumber\\
   & {\rm g}_x = \left(\frac{\lambdabar_c}{a_0}\right) ^\frac{3}{2} \frac{Q(R)}{R},
\end{align}
where, $P(R)$ and $Q(R)$ are the larger and smaller components of the electron Dirac radial wave function for the electron orbit $x$. They are calculated at the nuclear surface $R$. In our calculations, we have taken $R = 1.2 A^{1/3}$. The $P(R)$ and $Q(R)$ are solved by the Dirac bound-state radial equation using the subroutine RADIAL \cite{SALVAT1995151,SALVAT2019165} and our recent solver DiracSVT and its revised version using the finite difference method \cite{Mario1,Mario2}. The electron orbital $x$ is defined by $(N,k_x)$ (tabulated in \ref{tab:1}), where $N$ is the principal quantum number and the quantity $k_x$ (relative angular momentum quantum number) is related to the orbital and total angular momentum by 
$$k_x = (l-j)(2j+1),$$
\begin{align}\label{Eq.5}
        k_x &= l, \,\,\, \,\,\, \,\,\, \,\,\, \,\,\, \,\,\, \,\,\, \,\, j = l-1/2 \nonumber \\
        & = -(l+1),  \,\,\, j = l+1/2
\end{align}

\begin{table}[h]
    \centering
        \caption{The values of $k_x$ for different electron shells.}
    \label{tab:1}
    \begin{tabular}{cc|cc}
		\hline \hline
		Shell (x) & ($N$,$k_x$) & Shell (x) & ($N$,$k_x$)\\
		\hline
            K ($1s_{1/2}$) & (1,-1)  & N1 ($4s_{1/2}$) & (4,-1)\\
            
            L1 ($2s_{1/2}$) & (2,-1) & N2 ($4p_{1/2}$) & (4,+1)\\
            L2 ($2p_{1/2}$) & (2,+1) & N3 ($4p_{3/2}$) & (4,-2)\\
            L3($2p_{3/2}$) & (2,-2)  & N4 ($4d_{3/2}$) & (4,+2)\\
            
            M1 ($3s_{1/2}$) & (3,-1) & N5 ($4d_{5/2}$) & (4,-3)\\  
            M2 ($3p_{1/2}$) & (3,+1) & N6 ($4f_{5/2}$) & (4,+3)\\
            M3 ($3p_{3/2}$) & (3,-2) & N7 ($4f_{7/2}$) & (4,-4)\\
            M4 ($3d_{3/2}$) & (3,+2) & O1 ($5s_{1/2}$) & (5,-1)\\
            M5 ($3d_{5/2}$) & (3,-3) & O2 ($5p_{1/2}$) & (5,+1)\\

		\hline\hline
    \end{tabular}
\end{table}

As mentioned above, the $ft$-value is evaluated using the EC process for transitions in which $\beta^-$ decay is forbidden in neutral atoms. The EC half-life is given as
\begin{equation}\label{Eq.6}
    t_{1/2}^{\rm EC} = \kappa / \sum f_x C^{\rm EC}.
\end{equation}
Here, the phase-space function for the EC takes the same form as that used for $\beta_b$-decay \cite{Behrens1982ElectronRW,RevModPhys.49.77},
\begin{equation}\label{Eq.7}
    f_x = \frac{\pi}{2} \sigma_x \beta_x^2  B_x  q_x^2,
\end{equation}
here, $\sigma_x$ denotes the electron occupation number for a given shell. For closed shells,  $\sigma_x = 1$. We adopt the value for the  $\sigma_x$ as unity, similar to Ref. \cite{HARTMANN1992237}. The quantity $\beta_x$ is the bound-state electron radial wave function in the parent atom. $B_x$ represents the exchange and imperfect atomic overlap correction factor, which is considered as 1 in our calculations. The $q_x$ is the neutrino momentum, expressed as 
$$ q_x = Q_{\rm EC} - E^{\prime}_x.$$
Here, $Q_{\rm EC}$ is the atomic mass difference between the parent and daughter neutral atom, and $E^{\prime}_x$ denotes the binding energy of the captured electron in the daughter atom. The shape factor $C^{\rm EC}$ is related to the nuclear matrix elements. Since, the nuclear matrix elements depend only on the initial and final states, one can relate the shape factor for $\beta^-$ decay from an initial state to a final state with the shape factor of the reverse EC from final to initial state as follows
\begin{equation}\label{Eq.8}
    C_{i\rightarrow f}^{\beta^-} = \frac{2J_f+1}{2J_i+1} C_{f\rightarrow i}^{\rm EC} 
\end{equation}
Using the experimental half-life and evaluated $f_x$ of the EC process  from various atomic shells, we calculate the nuclear matrix elements for the $\beta^-$ decay, which are the used to determine the bound-state decay rate.

We now move to the study of $\beta^-$-decay to both the continuum- and bound-state for fully ionized atoms. We adopt the experimental partial half-life for the neutral atom $\beta^-$ decay and evaluate the $f_0t$ and $f_1t$ values for the allowed/first forbidden non-unique and first forbidden unique decays, respectively. These values are then employed to calculate the decay rates, as defined in Eq. \ref{Eq.1}, for fully ionized systems.
The $f_{0}$ has the form \cite{suhonen_2007}
\begin{equation}\label{Eq:9}
	f_{0}= \int_1^{w_0}pw_e(w_0-w_e)^2F_0(Z,w_e)dw_e.
\end{equation}

For the first forbidden unique transitions, $f_{1}$ \cite{suhonen_2007} can be written as
\begin{align}\label{Eq:10}
    f_{1}= &\int_1^{w_0} [(w_0-w_e)^2 + \lambda_{2} (w_e^2-1)] p \nonumber \\ & \times w_e(w_0-w_e)^2 F_0(Z,w_e)dw_e,
\end{align}
where $p = p_ec/m_ec^2 = \sqrt{w_e^2 -1}$ and $w_0 = Q_n /m_e c^2 + 1$ denote the momentum and the maximum energy of the emitted $\beta^-$ particle for a $Z-1 \rightarrow Z$ transition, respectively. The quantity 
\begin{equation}\label{Eq.11}
\lambda_{k_e}=\frac{F_{k_e-1}(Z,w_e)}{F_0(Z,w_e)},
\end{equation}
where, $F_{k_e-1}(Z,w_e)$ is the generalized Fermi function \cite{Behrens1982ElectronRW,PhysRevC.95.024327}, which takes the form
\begin{eqnarray} \label{Eq.12}
& F_{k_e-1}(Z,w_e) =4^{k_e-1}(2k_e)(k_e+\gamma_{k_e})[(2k_e-1)!!]^2 \nonumber \\
 & \times e^{\pi{y}}\left(\frac{2p_eR}{\hbar}\right)^{2(\gamma_{k_e}-k_e)}\left(\frac{|\Gamma(\gamma_{k_e}+iy)|}{\Gamma(1+2\gamma_{k_e})}\right)^2,
\end{eqnarray}
where the auxiliary quantities are defined as $\gamma_{k_e}=\sqrt{k_e^2-(\alpha{Z})^2}$ and $y=(\alpha{Zw_e}/p_ec)$.

The expression of the $f_m^*$ for the $\beta_b$-decay is given above in Eq. \ref{Eq.2}. The lepton phase space part for the continuum-state $\beta^-$ decay for allowed and first forbidden non-unique transitions is defined as
\begin{equation}\label{Eq.13}
	f_{m}^*= \int_1^{w_c}pw_e(w_c-w_e)^2F_0(Z,w_e)dw_e,
\end{equation}
and, for the first forbidden unique transitions, it is
\begin{align}\label{Eq.14}
    f_{m}^*= & \int_1^{w_0} [(w_c-w_e)^2 + \lambda_{2} (w_e^2-1)] p \nonumber \\
    & w_e(w_c-w_e)^2 F_0(Z,w_e)dw_e.
\end{align}

Here, $w_c$ is related to continuum-state decay energy $Q_c$ by $w_c = Q_c/m_ec^2 + 1$, and $Q_c = Q_n - \Delta B_e$.

    \begin{table*}[]
    \centering
    \caption{The selected $\beta_b$-decay candidates. The first two columns show the transitions from the initial state of the parent nucleus to the final state of the daughter nucleus, along with the spin-parity and excitation energies of both states. The quantity $Q_n$ represents the decay energy (in keV) for the neutral atom. The columns $Q_b(K)$ and $Q_b(L1)$ correspond to the $\beta_b$-decay energies for fully ionized atoms, where the emitted electron is created in the K- and L1-shell orbital, respectively.
    }
    \label{tab:2}
    \begin{tabular}{ccccccccc}
    \hline\hline
       Parent $\rightarrow$ daughter  & Transition & $Q_n$  & $B_n(Z)-B_n(Z-1)$ & $B_K$  & $Q_b(K)$  & $B_{L1}$  & $Q_b(L1)$  \\

        & [$E_i({\rm keV}), J_i^{\pi} \rightarrow E_f({\rm keV}), J_f^{\pi}$] & (keV) &  (keV) & (keV) & (keV) & (keV) & (keV) \\
       \hline
        $^{163}$Dy $\rightarrow$ $^{163}$Ho & [0.0, $\frac{5}{2}^- \rightarrow 0.0,  \frac{7}{2}^- $] & -2.863 \cite{Schweiger2024} & 12.5 & 65.137 & 49.774  & 15.746 & 0.383\\
        $^{187}$Re $\rightarrow$ $^{187}$Os & [0.0, $\frac{5}{2}^+ \rightarrow 0.0,  \frac{1}{2}^- $] & 2.5 & 15.2 & 85.614 & 72.914 & 20.921 & 8.221\\
        & $[0.0, \frac{5}{2}^+ \rightarrow 9.756,  \frac{3}{2}^- ]$ & -7.3 & 15.2 & 85.614 & 63.114 & 20.921 & -1.588\\
        $^{193}$Ir $\rightarrow$ $^{193}$Pt & [0.0, $\frac{3}{2}^+ \rightarrow 0.0,  \frac{1}{2}^- $] & -56.6 & 15.9 & 90.660 & 18.160 & 22.206 & -50.294\\

        $^{194}$Au $\rightarrow$ $^{194}$Hg & [0.0, $1^- \rightarrow 0.0,  0^+ $] & -27.9 & 16.8 & 95.898 & 51.198 & 23.544 & -21.156\\

        $^{202}$Tl $\rightarrow$ $^{202}$Pb & [0.0, $2^- \rightarrow 0.0, 0^+$] & -39.4 & 17.3 & 101.336 & 44.636 & 24.938 & -31.762\\
        
        $^{205}$Tl $\rightarrow$ $^{205}$Pb & [0.0,$\frac{1}{2}^+ \rightarrow 0.0, \frac{5}{2}^- $]    & -50.6 \cite{Leckenby2024} & 17.3 & 101.336 & 33.436 & 24.938 & -42.962\\
                                            & [0.0, $\frac{1}{2}^+ \rightarrow 2.329,\frac{1}{2}^-] $ & -52.9 & 17.3 & 101.336 & 31.136 & 24.938 & -45.262\\

        $^{215}$At $\rightarrow$ $^{215}$Rn & [0.0, $\frac{9}{2}^- \rightarrow 0.0,  \frac{9}{2}^+ $] & -88.0 & 18.7 & 112.842 & 6.142 & 27.903 & -78.797\\

        $^{222}$Rn $\rightarrow$ $^{222}$Fr & [0.0, $0^+ \rightarrow 0.0,  2^- $] & -6.0 & 19.2 & 115.858 & 90.658 & 28.683 & 3.483\\

        $^{227}$Ac $\rightarrow$ $^{227}$Th & $[0.0, \frac{3}{2}^- \rightarrow 0.0, (\frac{1}{2}^+) ]$ & 44.7 & 20.3 & 125.250 & 149.650 & 31.123 & 55.523\\
        & $[0.0, \frac{3}{2}^- \rightarrow 9.3, (\frac{5}{2}^+) ]$  & 35.4 & 20.3 & 125.250 & 140.350 & 31.123 & 46.223\\
        & $[0.0, \frac{3}{2}^- \rightarrow 24.5, (\frac{3}{2}^+) ]$  & 20.2 & 20.3 & 125.250 & 125.150 & 31.123 & 31.023\\
        & $[0.0, \frac{3}{2}^- \rightarrow 37.863, (\frac{3}{2}^-) ]$  & 6.8 & 20.3 & 125.250 & 111.750 & 31.123 & 17.623\\
        $^{228}$Ra $\rightarrow$ $^{228}$Ac & $[0.0, 0^+ \rightarrow 6.11, 1^- ]$ & 39.4 & 19.9 & 122.063 & 141.563 & 30.293 & 49.793\\
        & $[0.0, 0^+ \rightarrow 6.70, 1^+ ]$ & 38.8 & 19.9 & 122.063 & 140.963 & 30.293 & 49.193\\
        & $[0.0, 0^+ \rightarrow 20.6, 1^- ]$ & 24.9 & 19.9 & 122.063 & 127.063 & 30.293 & 35.293\\
        & $[0.0, 0^+ \rightarrow 33.07, 1^+ ]$ & 12.4 & 19.9 & 122.063 & 114.563 & 30.293 & 22.793\\

        $^{243}$Am $\rightarrow$ $^{243}$Cm & [0.0, $\frac{5}{2}^- \rightarrow 0.0,  \frac{5}{2}^+ $] & -6.9 & 24.0 & 145.740 & 114.840 & 36.493 & 5.593\\
        $^{246}$Bk $\rightarrow$ $^{246}$Cf & [0.0, $2^{(-)} \rightarrow 0.0,  0^+ $] & -120.2 & 23.0 & 153.124 & 9.924 & 38.444 & -104.756\\
        
         \hline
         \hline
    \end{tabular}
\end{table*}
\section{Results and Discussion}\label{Results}
In the present study, we have investigated the $\beta_b$-decay rates of fully ionized atoms. The selected candidate nuclei are listed in Table \ref{tab:2}, along with their decay energies in both the neutral and bare atoms. In these transitions, the emitted electron can occupy the $1s_{1/2}$ and $2s_{1/2}$ orbital of the daughter atom. There are 11 transitions with negative $Q_n$-values, indicating that $\beta^-$ decay is energetically not allowed in neutral atoms but becomes possible in bare atoms. As shown in the table, the $\beta_b$-decay transitions of $^{187}$Re (to the first excited state), $^{193}$Ir, $^{194}$Au, $^{202}$Tl, $^{205}$Tl, $^{215}$At, and $^{246}$Bk are only permitted when the emitted electron is created in the K-shell. In these cases, we extract the nuclear matrix elements for the $\beta_b$-decay using the experimental EC half-lives. Among them, EC half-lives have been experimentally measured for 7 transitions, which are presented in Table  \ref{tab:3}.

The first experimental observation of $\beta_b$ decay was reported by Jung \textit{et al.} in 1992 for fully ionized $^{163}$Dy, using the experimental storage ring (ESR) facility at GSI \cite{PhysRevLett.69.2164}. The measured half-life was estimated to be 47(+5-4) days. This was followed by measurements of the $\beta_b$-decay for $^{187}$Re in 1996 \cite{PhysRevLett.77.5190} and $^{207}$Tl in 2005 \cite{PhysRevLett.95.052501}. More recently, a new $\beta_b$-decay experiment for bare $^{205}$Tl was carried out and a half-life of 291(+33-27) days was observed \cite{Leckenby2024,PhysRevLett.133.232701,EurPhysJA.61.130,leckenby2025bayesianmontecarloapproaches}.

Following the Takahashi-Yokoi model, several theoretical efforts have been made to calculate the decay rates of the $\beta_b$-decay for bare atoms. Xiao \textit{et al.} have explored the $\beta_b$-decay half-lives of experimentally measured highly ionized atoms such as $^{163}$Dy, $^{187}$Re, and $^{205}$Tl \cite{PhysRevC.110.054308}, as well as the stellar beta decay rate of $^{204}$Tl \cite{PhysRevC.110.015806}, employing the projected shell model \cite{S0218301395000250,SUN1996375,TAN2020135432,CHEN2024138338,WANG2024138515}. Furthermore, allowed $\beta^-$ decay rates for bare atoms in the mass range $A = 59$–$81$ have been computed using the nuclear shell model, as reported in Ref. \cite{PhysRevC.108.015805}.

\begin{table*}
    \centering
    \caption{Experimental EC half-lives and log$ft$ values for the candidates where available \cite{NNDC}}
    \label{tab:3}
    \begin{tabular}{ccccccccc}
    \hline\hline
       Parent $\rightarrow$ daughter  & Transition & $Q_{\rm EC}$  & $t^{\rm EC}_{1/2}$ & log$ft$(EC) \\

        & [$E_i({\rm keV}), J_i^{\pi} \rightarrow E_f({\rm keV}), J_f^{\pi}$] & (keV) &   & \\
       \hline
        $^{163}$Ho $\rightarrow$ $^{163}$Dy & [0.0, $\frac{7}{2}^- \rightarrow 0.0,  \frac{5}{2}^- $] & 2.863 \cite{Schweiger2024} & 4570(25) y  & 4.91(1)\\

        $^{193}$Pt $\rightarrow$ $^{193}$Ir & [0.0, $\frac{1}{2}^- \rightarrow 0.0,  \frac{3}{2}^+ $] & 56.6 & 50(6) y & 7.16(6)\\

        $^{194}$Hg $\rightarrow$ $^{194}$Au & [0.0, $0^+ \rightarrow 0.0,  1^- $] & 27.9 & 447(52) y  & 8.4(3)\\

        $^{202}$Pb $\rightarrow$ $^{202}$Tl & [0.0, $0^+ \rightarrow 0.0, 2^-$] & 39.4 & 5.25 (28) $\times 10^4$ y & 8.95(13)$^{\rm 1u}$\\
        
        $^{205}$Pb $\rightarrow$ $^{205}$Tl & [0.0,$\frac{5}{2}^- \rightarrow 0.0, \frac{1}{2}^+ $]    & 50.6 \cite{Leckenby2024} & 17.0(9) $\times 10^6$ y & 11.73(3)$^{\rm 1u}$  \\
                                            & [2.329, $\frac{1}{2}^- \rightarrow 0.0,\frac{1}{2}^+] $ & 52.9 & 793 d \cite{Leckenby2024}\\

        $^{243}$Cm $\rightarrow$ $^{243}$Am & [0.0, $\frac{5}{2}^+ \rightarrow 0.0,  \frac{5}{2}^- $] & 6.9 & 29.1(1) y & $\leq 7.2$\\
        
         \hline
         \hline
    \end{tabular}
\end{table*}

In this work, we have calculated the half-lives and decay rates for bare atoms using the latest AME2020 mass evaluation data and the formalism outlined in Section \ref{Methodology}, presented in Tables \ref{tab:4}-\ref{tab:5}. In Ref. \cite{YASUMI1994229}, the EC decay rates involving the M1 and M2 shells were measured, yielding a log$ft$ value of 4.993 and a $Q_n$ value of 2.710 keV. This log$ft$ value was later used to estimate the $\beta_b$-decay half life in Ref. \cite{PhysRevC.104.024304}. According to the updated AME2020 data \cite{Wang_2021}, the $Q_n$ value has been revised to 2.9 keV. Recently, high-precision Penning-trap measurements reported a $Q_n$ value of 2.863 keV for the EC decay of $^{163}$Ho \cite{Schweiger2024}, which we have adopted in our calculations. It should be mentioned that in Ref. \cite{PhysRevLett.69.2164}, the $\beta_b$-decay rate of $^{163}$Dy was measured to be $(1.72 \pm 0.1 \pm 0.07) \times 10^{-7} {\rm s}^{-1}$. In that paper, they also evaluated the ratio of decay rates between the $\beta_b$-decay and EC, however, they only considered the $\beta_b$-decay of $^{163}$Dy into the K-shell and EC from the M1 and M2 shells of $^{163}$Ho. They also included a factor of $1/2$ to account for the different number of electrons participating in $\beta_b$-decay and EC process. It may be useful to mention that they used the electron radial wave function [see Eq. (8) therein] of the daughter atom for the $\lambda_{\rm EC}$ instead of the parent atom, which may not be proper. 

In our calculations, we have used the binding energies from Ref. \cite{RevModPhys.39.125,BE}. It should be pointed out that in Ref. \cite{PhysRevC.104.024304}, the EC decay energy was defined as the difference in atomic mass between the parent and daughter atoms, minus the electron binding energy of the respective shell (K, L, M, ...) in the parent atom, instead of from the daughter atom. Furthermore, it was taken for the neutral atom rather than a fully ionized atom. Those calculations may not be fully correct. In our case, when considering the binding energies of the neutral $^{163}$Dy atom, it becomes clear that EC from the M1, M2, N1, N2, and higher shells is energetically feasible.

In the calculation of the phase space factor for the EC for $^{163}$Ho $\rightarrow$ $^{163}$Dy transition, we first consider an electron capture from the M1 orbital. Using this, we obtained a log$ft$ value of 4.612. Based on the methodology described above, we determineed the nuclear matrix elements for the $\beta_b$-decay, where the $ft$ value is given by  $\kappa/C^{\beta^-}$. The value of the decay constant $\kappa$ is taken to be 6144.5 s, as reported in Ref. \cite{Leckenby2024}.  The calculated half-life is 15.60 days, and the corresponding decay rate is $5.14 \times 10^{-7}$ s$^{-1}$. We further extended our calculation by including EC contributions from additional atomic orbitals. When EC from the entire M-shell is considered, the resulting half-life for the $\beta_b$-decay increases to 17.11 days. Including contributions from both M- and N-shells leads to a further increase in half-life, reaching 79.94 days. Finally, our calculated half-life comes out to be 121.70 days for the fully ionized atom, which is in reasonable agreement with the experimental value. 

For the transition from the ground state of $^{205}$Tl to the first excited state of $^{205}$Pb, we obtained a half-life of 412.98 days, which is close to the measured value of 291(+33-27) days. In the calculation of Ref. \cite{FREEDMAN1988267}, a factor of $1/2$ was introduced to reflect the presence of the single 1$s$ electron produced in the $\beta_b$-decay of $^{205}$Tl, in contrast to two available 2$s$ electrons for EC. In their analysis, the wavefunctions used to evaluate the decay rates of $\beta_b$-decay and EC correspond to the highly ionized Pb$^{81+}$ ion final state with a bound 1$s$ electron, and a neutral lead atom initial state from which a 2$s$ electron is captured, respectively.

\begin{widetext}

\begin{longtable}{ccccccc}
\caption{The calculated half-lives and decay rates of fully ionized atoms. The spin-parity of parent and daughter nuclei with their decay mode, as well as the experimental half-life are shown in column I. The second column provides information about the EC from different atomic orbitals. The log$ft$ values for the EC and $\beta_b$-decay are given in columns III and V, respectively. The calculated $f_m^*$ for the $\beta_b$-decay is listed in column IV. }\\
    \hline\hline

       Parent $\rightarrow$ daughter  &    Orbital & log$ft$(EC) & $f_m^*$  & log$ft$(${\beta_b}$) & $\lambda_{\beta_b}$ (s$^{-1}$) & $t_{1/2}^{\beta_b}$\T\B\\
       \hline
       $^{163}$Dy $\rightarrow$ $^{163}$Ho& \T\\      
        $[0.0, \frac{5}{2}^- \rightarrow  0.0, \frac{7}{2}^-] $ & M1 & 4.612 &  2.279$\times 10^{-2}$ & 4.488  &  $5.14 \times 10^{-7}$ & 15.60 d\\      
  allowed    &  M1-2& 4.652 &  &  4.527   & $4.69 \times 10^{-7}$ & 17.11\\
             &  M1-5& 4.652 &  & 4.527    & $4.69 \times 10^{-7}$& 17.11\\
            $47^{+5}_{-4}$ d &  M,N1& 5.299 &  & 5.174   & $1.06 \times 10^{-7}$& 75.88  \\
             &  M,N1-2& 5.322 &  & 5.197  & $1.00 \times 10^{-7}$& 79.94\\
             &  M,N1-5& 5.322 &  & 5.197  & $1.00 \times 10^{-7}$& 79.94\\
             &  M,N1-5,O1& 5.495 &  & 5.370 & $6.74 \times 10^{-8}$ & 119.05\\
             &  M,N1-5,O1-2& 5.505 &  & 5.380  & $6.59 \times 10^{-8}$& 121.70\B\\
             
        \hline
       $^{205}$Tl $\rightarrow$ $^{205}$Pb& \T\\      
        $[0.0,\frac{1}{2}^+ \rightarrow 0.0, \frac{5}{2}^- ]$ & L1 & 12.524 & 1.385$\times 10^{-4}$ & 12.047   & $8.61 \times 10^{-17}$ & 2.55 $\times 10^8$y\\      
  unique    &  L1-2& 12.561  &  &  12.084   & $7.91 \times 10^{-17}$& 2.78 $\times 10^8$\\
             &  L1-3& 12.561 &  &  12.084  & $7.91 \times 10^{-17}$ & 2.78 $\times 10^8$\\        
             &  L,M1& 12.734 &  & 12.256  & $5.32 \times 10^{-17}$& 4.13 $\times 10^8$ \\
             &  L,M1-2& 12.748 &  & 12.271  & $5.15 \times 10^{-17}$& 4.27 $\times 10^8$ \\ 
             &  L,M1-5& 12.748 &  & 12.271  & $5.15 \times 10^{-17}$& 4.27 $\times 10^8$ \\
             &  L,M,N1& 12.808 &  & 12.331  & $4.48 \times 10^{-17}$& 4.91 $\times 10^8$ \\
             &  L,M,N1-2& 12.814 &  & 12.337 & $4.42 \times 10^{-17}$& 4.97 $\times 10^8$ \\
             &  L,M,N1-2,O1& 12.842 &  & 12.365  & $4.14 \times 10^{-17}$& 5.30 $\times 10^8$\\
             &  L,M,N1-2,O1-2& 12.845 &  & 12.368  & $4.12 \times 10^{-17}$& 5.34 $\times 10^8$\B\\
             \\
                $[0.0,\frac{1}{2}^+ \rightarrow 2.329, \frac{1}{2}^- ]$ & L1 & 5.686 & 2.798$\times 10^{-2}$ & 5.686   & $4.00 \times 10^{-8}$ & 200.54 d \T\\      
  non-unique    &  L1-2& 5.722 &  &  5.722  & $3.68 \times 10^{-8}$& 218.22 \\
             &  L1-3& 5.722 &  &  5.722   & $3.68 \times 10^{-8}$& 218.22\\        
             &  L,M1& 5.891 &  & 5.891    & $2.50 \times 10^{-8}$ & 321.49 \\
         291$^{+33}_{-27}$ d    &  L,M1-2& 5.904 &  & 5.904  & $2.42 \times 10^{-8}$& 331.97 \\ 
             &  L,M1-5& 5.904 &  & 5.904   & $2.42 \times 10^{-8}$& 331.97\\
             &  L,M,N1& 5.963 &  & 5.963  &  $2.11 \times 10^{-8}$& 380.16\\
             &  L,M,N1-2& 5.969 &  & 5.969   & $2.08 \times 10^{-8}$& 385.26\\
             &  L,M,N1-2,O1& 5.996 &  & 5.996   & $1.96 \times 10^{-8}$& 410.28\\
             &  L,M,N1-2,O1-2& 5.999 &  & 5.999   & $1.94 \times 10^{-8}$& 412.98\B\\
       \hline          
        $^{193}$Ir $\rightarrow$ $^{193}$Pt & \T\\
        $[0.0, \frac{3}{2}^+ \rightarrow 0.0,  \frac{1}{2}^- ]$ & L1 &  7.023  & 7.037$\times 10^{-3}$ & 7.324  & $2.32 \times 10^{-10}$& 94.92 y\\
    non-unique  &  L1-2& 7.055 && 7.356  & $2.15 \times 10^{-10}$& 102.25\\
                &  L1-2,M1& 7.209 && 7.510  & $1.51 \times 10^{-10}$& 145.25\\
                &  L1-2,M1-2& 7.220 && 7.521  & $1.47 \times 10^{-10}$& 149.72\\
                &  L1-2,M1-2,N1& 7.275 && 7.576  & $1.30 \times 10^{-10}$& 169.65\\
                &  L1-2,M1-2,N1-2& 7.279 && 7.580  & $1.28 \times 10^{-10}$& 171.51\\
                &  L1-2,M1-2,N1-2,O1& 7.305 && 7.606  & $1.21 \times 10^{-10}$& 181.81\\
                &  L1-2,M1-2,N1-2,O1-2& 7.307 && 7.608  & $1.20 \times 10^{-10}$& 182.79\B \\
                \hline
                $^{194}$Au $\rightarrow$ $^{194}$Hg  & \T \\
        $[0.0, 1^- \rightarrow 0.0,  0^+ ]$ & L1 &  7.041  & 6.534$\times 10^{-2}$ & 7.519  &  $1.37 \times 10^{-9}$& 16.02 y\\
    non-unique  &  L1-2& 7.078 && 7.555   & $1.26 \times 10^{-9}$& 17.42\\
                &  L1-2,M1& 7.357 && 7.834   & $6.64 \times 10^{-10}$& 33.11\\
                &  L1-2,M1-2& 7.376 && 7.853  & $6.35 \times 10^{-10}$& 34.62\\
                &  L1-2,M1-2,N1& 7.467 && 7.944  & $5.15 \times 10^{-10}$& 42.67\\
                &  L1-2,M1-2,N1-2& 7.475 && 7.952 & $5.06 \times 10^{-10}$ & 43.48\\
                &  L1-2,M1-2,N1-2,O1& 7.516 && 7.993  & $4.60 \times 10^{-10}$& 47.75\\
                &  L1-2,M1-2,N1-2,O1-2& 7.520 && 7.997  & $4.56 \times 10^{-10}$& 48.18\B\\
                \hline
$^{202}$Tl $\rightarrow$ $^{202}$Pb& \T\\      
        $[0.0,2^- \rightarrow 0.0, 0^+ ]$ & L1 & 9.683 & 4.407$\times 10^{-4}$ & 10.382   &$1.27 \times 10^{-14}$& 1.73 $\times 10^6$ y \\      
  unique    &  L1-2& 9.720 &  &  10.419   & $1.16 \times 10^{-14}$& 1.89 $\times 10^6$\\        
             &  L1-2,M1& 9.926 &  & 10.625   & $7.25 \times 10^{-15}$& 303 $\times 10^6$\\
             &  L1-2,M1-2& 9.942 &  & 10.641   & $6.98 \times 10^{-15}$& 3.15 $\times 10^6$\\ 
             &  L1-2,M1-2,N1& 10.013 &  & 10.712  & $5.93 \times 10^{-15}$& 3.71 $\times 10^6$\\
             &  L1-2,M1-2,N1-2&  10.020 &  & 10.719  & $5.84 \times 10^{-15}$& 3.77 $\times 10^6$\\
             &  L1-2,M1-2,N1-2,O1& 10.052 &  & 10.751  & $5.42 \times 10^{-15}$& 4.06 $\times 10^6$\\
             &  L1-2,M1-2,N1-2,O1-2&  10.056 &  & 10.755  & $5.37 \times 10^{-15}$& 4.09 $\times 10^6$\B\\
             \hline
        $^{243}$Am $\rightarrow$ $^{243}$Cm & \T \\
        $[0.0, \frac{5}{2}^- \rightarrow 0.0,  \frac{5}{2}^+]$ & M1 & 5.991 & 1.1333 & 5.991  & $8.01 \times 10^{-7}$& 10.01 d\\
    non-unique & M1-2 & 6.124 && 6.124  & $5.90 \times 10^{-7}$& 13.59\\
               & M1-2,N1 & 7.299 && 7.299  & $3.94 \times 10^{-8}$& 203.49\\
               & M1-2,N1-2 & 7.365 && 7.364  & $3.39 \times 10^{-8}$& 236.44\B\\
             \hline
         \hline
    
    \label{tab:4}
\end{longtable}

\begin{table}[h]
    \centering
    \caption{The calculated ratio of $\beta_b$-decay for bare atoms and EC for neutral atoms.}
    \label{tab:5}
    \begin{tabular}{ccccccccc}
    \hline\hline
    
       Parent $\rightarrow$ daughter  &    Orbital& $f_m^*$  & $\lambda_{\beta_b}$/$\lambda_{{\rm EC}}$ \\
       \hline
        $^{187}$Re $\rightarrow$ $^{187}$Os & \\
        $[0.0, \frac{5}{2}^{+} \rightarrow 9.756,  \frac{3}{2}^-]$ & M1 & 7.301$\times 10^{-2}$  & $2.81 \times 10^{3}$ \\
    non-unique &  M1-2&& $2.57 \times 10^{3}$ \\
    &  M1-2,N1&& $1.36 \times 10^{3}$ \\
 32.9$\pm2.0$ y   &  M1-2,N1-2&& $1.30 \times 10^{3}$ \\
    &  M1-2,N1-2,O1&& $1.03 \times 10^{3}$ \\
    &  M1-2,N1-2,O1-2&& $1.01 \times 10^{3}$ \\
        \hline
        $^{215}$At $\rightarrow$ $^{215}$Rn & \\
        $[0.0, \frac{9}{2}^- \rightarrow 0.0,  \frac{9}{2}^+]$ & L1 & 1.484$\times10^{-3}$  & $4.23 \times 10^{-2}$  \\
    non-unique &  L1-2&& $3.85 \times 10^{-2}$ \\
                &  L1-2,M1&& $2.78 \times 10^{-2}$ \\
                &  L1-2,M1-2&&  $2.69 \times 10^{-2}$ \\
                &  L1-2,M1-2,N1&&  $2.40 \times 10^{-2}$ \\
                &  L1-2,M1-2,N1-2&&  $2.37 \times 10^{-2}$ \\
                &  L1-2,M1-2,N1-2,O1&&  $2.25 \times 10^{-2}$ \\
                &  L1-2,M1-2,N1-2,O1-2&&  $2.23 \times 10^{-2}$ \\
                
        \hline
        $^{222}$Rn $\rightarrow$ $^{222}$Fr & \\
        $[0.0, 0^+ \rightarrow 0.0,  2^-]$ & M1 & 1.096$\times 10^{-2}$  & $1.03 \times 10^{4}$ \\
    unique &  M1-2&& $8.81 \times 10^{3}$  \\
                &  M1-2,N1&& $1.87 \times 10^{3}$  \\
                &  M1-2,N1-2&&  $1.70 \times 10^{3}$  \\
                &  M1-2,N1-2,O1&&  $1.13 \times 10^{3}$  \\
                &  M1-2,N1-2,O1-2&&  $1.08 \times 10^{3}$  \\
        \hline
        $^{246}$Bk $\rightarrow$ $^{246}$Cf & \\
        $[0.0, 2^{(-)} \rightarrow 0.0,  0^+]$ & L1 & 3.755$\times 10^{-6}$  & $4.09 \times 10^{-6}$  \\
    unique &  L1-2&& $3.58 \times 10^{-6}$ \\
                &  L1-2,M1&& $3.38 \times 10^{-6}$ \\
                &  L1-2,M1-2&&  $2.94 \times 10^{-6}$ \\
                &  L1-2,M1-2,N1&&  $2.87 \times 10^{-6}$ \\
                &  L1-2,M1-2,N1-2&&  $2.81 \times 10^{-6}$ \\
                &  L1-2,M1-2,N1-2,O1&&  $2.63 \times 10^{-6}$ \\
        
             \hline
         \hline
    \end{tabular}
\end{table}

\begin{table}
    \centering
    \caption{The calculated $\beta_b$-decay half-lives and beta decay rates of fully ionized atoms. All the decay rates are in units of `${\rm s}^{-1}$'. The decay type distinguishes between allowed, first forbidden unique, and first forbidden non-unique transitions. The log$ft$ value and decay rate for the neutral atom are presented in columns III and IV, respectively. The bound and continuum state beta-decay rates for bare atoms are listed in columns V and VI, respectively. Columns VII and VIII show the ratios $\lambda_{\beta_b}$/$\lambda_{\beta_c}$ and $\lambda_{bare}$/$\lambda_{\rm neutral}$ for all the transitions. In the last column, the half-lives of the bare atoms are predicted.}
     \label{Tab:6}
    \begin{tabular}{cccccccccc}
    \hline\hline
    
       Parent $\rightarrow$ daughter & Decay type &log$ft$($\beta^-$)  & $\lambda_{\rm neutral}$  & $\lambda_{\beta_b}$ & $\lambda_{\beta_c}$ & $\lambda_{\beta_b}$/$\lambda_{\beta_c}$ & $\lambda_{\rm bare}$/$\lambda_{\rm neutral}$  & $t_{1/2}^{\rm bare}$\\
       \hline
        $^{187}$Re $\rightarrow$ $^{187}$Os &  \T\\
        $[0.0, \frac{5}{2}^+ \rightarrow 0.0,  \frac{1}{2}^-]$ & unique & 11.224 &5.08$\times10^{-19}$& 8.22$\times10^{-15}$ & 0 & & 1.62$\times10^{4}$ & 2.67 $\times 10^{6}$ y\B\\
\hline
        $^{227}$Ac $\rightarrow$ $^{227}$Th & \T\\
        $[0.0, \frac{3}{2}^- \rightarrow 0.0, (\frac{1}{2}^+) ]$ & non-unique  & 7.164 &5.45$\times10^{-10}$& 5.71$\times10^{-8}$ & 8.74$\times10^{-11}$ & 6.53$\times10^{2}$ & 1.05$\times10^{2}$\\
        $[0.0, \frac{3}{2}^- \rightarrow 9.3, (\frac{5}{2}^+) ]$  & non-unique & 7.045 &3.53$\times10^{-10}$& 6.59$\times10^{-8}$ & 2.70$\times10^{-11}$ & 2.44$\times10^{3}$ & 1.87$\times10^{2}$ & 31.31 d\\
        $[0.0, \frac{3}{2}^- \rightarrow 24.5, (\frac{3}{2}^+) ]$ & non-unique & 6.854 &1.01$\times10^{-10}$& 8.15$\times10^{-8}$ & 0 &  & 8.07$\times10^{2}$\\
        $[0.0, \frac{3}{2}^- \rightarrow 37.863, (\frac{3}{2}^-) ]$  & allowed   & 6.954 &3.03$\times10^{-12}$& 5.16$\times10^{-8}$ & 0 &  & 1.70$\times10^{4}$\B\\
        \hline
        $^{228}$Ra $\rightarrow$ $^{228}$Ac & \T\\
        $[0.0, 0^+ \rightarrow 6.11, 1^- ]$ & non-unique & 7.122 &3.82$\times10^{-10}$& 5.18$\times10^{-8}$ & 4.57$\times10^{-11}$ & 1.13$\times10^{3}$ & 1.36$\times10^{2}$ \\
         $[0.0, 0^+ \rightarrow 6.70, 1^+ ]$ & allowed   & 6.500 &1.53$\times10^{-9}$& 2.15$\times10^{-7}$ & 1.74$\times10^{-10}$ & 1.26$\times10^{3}$ & 1.41$\times10^{2}$ & 2.04 d\\
         $[0.0, 0^+ \rightarrow 20.6, 1^- ]$ & non-unique& 6.219 &7.65$\times10^{-10}$& 3.34$\times10^{-7}$ & 6.10$\times10^{-12}$ & 5.47$\times10^{4}$ & 4.37$\times10^{2}$\\
         $[0.0, 0^+ \rightarrow 33.07, 1^+ ]$ & allowed  & 5.131 &1.15$\times10^{-9}$&3.33$\times10^{-6}$ & 0 & &2.90$\times10^{3}$& \B\\

             \hline
         \hline
    \end{tabular}
\end{table}

\end{widetext}

From the analysis of the decays $^{163}$Dy $\rightarrow$ $^{163}$Ho and $^{205}$Tl $\rightarrow$ $^{205}$Pb, it is evident that the dominant contributions to the $\beta_b$-decay originate from the $s_{1/2}$ and $p_{1/2}$ orbitals. Therefore, for the remaining transitions considered in this study, we have tabulated contributions from these orbitals. There is no available experimental data for the decay of bare $^{193}$Ir, $^{194}$Au, $^{202}$Tl, and $^{243}$Am atoms. Our calculations show that the half-lives of $^{193}$Ir, $^{194}$Au, and $^{202}$Tl are on the order of years, while that of $^{243}$Am is 236.44 days, suggesting it could be a promising candidate for future experimental investigation. For the decays of $^{215}$At, $^{222}$Rn, $^{246}$Bk, and $^{187}$Re (to the first excited state), EC has not yet been measured experimentally. Consequently, for these cases, we have reported the ratio of the $\beta_b$-decay rate to the EC rate in Table \ref{tab:5}.

Additionally, we have studied the decay rates for transitions that are energetically possible in both neutral and bare atoms. For such transitions, the experimental half-lives and branching ratios of $\beta^-$-decay \cite{NNDC} are employed to compute the rates in neutral atoms ($\lambda_{\rm neutral}$). The $f_0t({\rm or} \,\, f_1t)$ values, using the formulas expressed in Eqs. \ref{Eq:9}-\ref{Eq:10}, are then determined depending on the type of the transition. Using these values, we calculate the decay rate for both continuum- and bound-state $\beta^-$ decay, with the results summarized in Table \ref{Tab:6}. It is evident from the Table \ref{Tab:6} that the continuum-state decay rate ($\lambda_{\beta_c}$) is much smaller as compared to the bound-state decay rate ($\lambda_{\beta_b}$). In particular, for the decay of $^{187}$Re, the continuum-state $\beta^-$ decay rate is 0, which means the only possible mode is the $\beta_b$-decay. In Table \ref{Tab:6}, we define the total decay rate for fully ionized atoms as $\lambda_{\rm bare} = \lambda_{\beta_b} + \lambda_{\beta_c}$ and the corresponding total half-life as ln 2$/\sum_i(\lambda_{\rm bare})_i$, where index $i$ runs over all possible decay types. The ratio of $\lambda_{\rm bare}/\lambda_{\rm neutral}$ reveals a significant enhancement in decay rate by a factor of $10^2-10^4$ in the case of bare atoms.   

\section{Summary}\label{Summary}
To summarize, we have done a systematic study on the $\beta_b$-decay of fully ionized atoms where the nuclear matrix elements (the $ft$ values) used to determine the $\beta_b$-decay half-life are taken from the comparative EC and $\beta$ decay processes as in the Takahashi-Yokoi model and the impact of the capture to different electron orbitals are investigated thoroughly. Firstly, we have considered only $\beta_b$-decay from bare atoms for which normal $\beta$ decay is forbidden and have evaluated their half-lives based on the time-mirrored EC process. The calculated half-lives for fully ionized $^{163}$Dy and $^{205}$Tl atoms are predicted to be 121.70 and 412.98 days, respectively, which are in good agreement with available experimental data. Further, we have done a survey on some $\beta^-$ decay transitions which are possible in both neutral and bare cases, and presented those cases where the ratio of their decay rates is significant and is in the range of $10^{2-4}$. 

\section{ACKNOWLEDGMENTS}
We thank the financial support from the Olle Engkvist Foundation and the computational resources provided by the National Academic
Infrastructure for Supercomputing in Sweden (NAISS) at PDC, KTH.
\bibliography{ref}
\end{document}